\documentclass[12pt]{article}

\usepackage{color}
\usepackage{epsfig,
amssymb,
amsfonts,
amsmath,
graphicx,
dsfont,
xfrac,
authblk,
subcaption
}

\definecolor{mygray}{gray}{0.5}

\usepackage[normalem]{ulem}

\usepackage{cite}
\usepackage[colorlinks=true,linkcolor=blue,citecolor=red]{hyperref}

\newcommand{\be}{\begin{equation}}
\newcommand{\ee}{\end{equation}}
\newcommand{\bea}{\begin{eqnarray}}
\newcommand{\eea}{\end{eqnarray}}

\title{Balanced Gain-and-Loss Optical Waveguides: Exact Solutions for Guided Modes in Susy-QM}

\author[${1}$]{Sara Cruz~y~Cruz}
\author[${2}$]{Alejandro Romero-Osnaya}
\author[${2}$]{Oscar Rosas-Ortiz}

\affil[${1}$]{\footnotesize Instituto Polit\'ecnico Nacional, UPIITA, Av I.P.N 2580,  M\'exico City C.P. 07340, Mexico}

\affil[${2}$]{\footnotesize Physics Department, Cinvestav, AP 14-740, 07000
M\'exico City, Mexico}

\date{}
\begin{document}

\maketitle

\begin{abstract}
The construction of exactly solvable refractive indices allowing guided TE modes in optical waveguides is investigated within the formalism of Darboux-Crum transformations. We apply the finite-difference algorithm for higher-order supersymmetric quantum mechanics to obtain complex-valued refractive indices admitting all-real eigenvalues in their point spectrum. The new refractive indices are such that  their imaginary part gives zero if it is integrated over the entire domain of definition. This property, called condition of zero total area, ensures the conservation of optical power so the refractive index shows balanced gain and loss. Consequently, the complex-valued refractive indices reported in this work include but are not limited to the parity-time invariant case. 
\end{abstract}


\section{Introduction}

Optical waveguides with arbitrarily shaped index profiles have been the subject of intensive research over recent decades. The accelerated development of fiber optics technologies and optical communications has prompted new ways of visualizing the interaction of light with matter. In addition, advances in material and optical technologies open wider possibilities to design index profiles on demand, so more sophisticated waveguiding-based optical devices may be achieved in a short time.

The behavior of light beams in spatially inhomogeneous media of refractive index $n$ is formally studied through the wave equation derived from the Maxwell equations~\cite{Oko82,Sny83,Oka06,Cal07}. As the pulse spread at the receiving end of a waveguide can be reduced for refractive indices of quadratic (parabolic) profile, it is expected that shaping the profile more freely we might be able to reduce further this undesired phenomenon~\cite{Oko82}. Such expectation motivates the investigation of arbitrarily shaped index profiles addressed to optical waveguiding. Nevertheless, analytical solutions of the equation that rules the propagation of light in this kind of media exist only for limited refractive index profiles. The WKB, Rayleigh-Ritz, power-series expansion, finite element or perturbation methods are then applied to obtain some approximated solutions~\cite{Oko82,Sny83,Oka06,Cal07}, even numerical integration is usable. In this respect, the transformation introduced by Darboux in 1882 to study Sturm-Liouville problems deserves special attention.

The Darboux method permits the intertwining of two differential equations in such a form that their solutions are connected through a simple differential relationship; further improvements are due to Crum~\cite{Mie04,Rog02}. In the 1970s, such a method received a lot of interest in soliton theory since it is linked to the B\"acklund transformation, used to study the soliton-like solutions of diverse nonlinear differential equations~\cite{Rog02}. The Darboux approach attracted renewed attention in the 1980s, after the inauguration of supersymmetric quantum mechanics by Witten~\cite{Wit81}, because it is also in the core of the factorization method introduced by Dirac `as a little stratagem to solve the spectral problem for the one-dimensional quantum oscillator'~\cite{Mie04}. Diverse supersymmetric formulations were immediately addressed to the spectral design of exactly solvable models in quantum mechanics~\cite{Mie84, And84a,And84b,And84c,Nie84}.

Supersymmetric quantum mechanics (Susy-QM) is presently a very robust formulation embracing a very wide set of applications~\cite{Mie04,Bag00,Coo01}, including optics~\cite{Chu94,Raz19,May16,Mil92,Bag90,Dia99,Mie00,Mir13a,Mir14,Hei14,Mac18}, where supersymmetry may be interpreted as describing two light beams of different colors that form a standing periodic interference pattern along the waveguide axis~\cite{Chu94}. In the Helmholtz regime, the supersymmetric partners can be constructed to display parabolic~\cite{Raz19} or sech-like index profile~\cite{Mil92,Dia99,Mie00,May16}. They also show identical coefficients of transmission and reflection for any angle of incidence~\cite{Mie04,Bag90,Dia99,Mie00,Mir13a}, which may render them perfectly indistinguishable to an external observer~\cite{Mir14}. The impressing breakthrough of supersymmetry in optics is that superpartner configurations are experimentally realized in coupled optical networks, where the light dynamics is directly observed~\cite{Hei14} and the conditions for degeneracy breaking can be studied~\cite{Mac18}. The theoretical modeling of diverse refractive indices using the powerful Darboux–Crum (supersymmetric) formulation~\cite{Mie04,Mie84,And84a,And84b,And84c,Nie84,Mie00,Bag99} is therefore interesting by itself.

The Darboux-deformation of Hermitian systems is just one of the options permitted by the method. Indeed, non-Hermitian structures may be also developed~\cite{Bay96,Can98,And99,Bag01,Fer08}, where some initial insights connecting supersymmetry with parity-time (PT) symmetry~\cite{Bag02} shown additional applications of the supersymmetric approach. In this context, important theoretical achievements within the PT-symmetric formulation~\cite{Gan07} envisioned the observation of parity-time symmetry in optical laboratories~\cite{Rut10}, so the practical implementation of Susy-QM could also be achieved in optics. Moreover, as the reality of the spectrum is not granted a priori for non-Hermitian models, the demonstration that PT symmetry implies real spectrum~\cite{Ben98} stimulated the systematic search of PT-symmetric systems in quantum mechanics~\cite{Lev20}. Nevertheless, further improvements shown that $PT$-symmetry is not a necessary condition for the spectrum reality~\cite{Mos02,Mos10}, a fact confirmed in diverse supersymmetric models of non-Hermiticity~\cite{Ros15,Ros18,Bla19,Bag19,Zel20a,Zel20d}, where the imaginary part of a wide class of complex-valued potentials lead to balanced gain and loss probability without the necessity of PT symmetry. The latter property opens new possibilities in optical design, where the gain-loss manipulation includes but is not limited to PT symmetry~\cite{Gbu18}. 

In the present work we generate graded refractive indices with both real- and complex-valued profiles. Our interest is addressed to the study of guided TE modes propagating in media with balanced gain-and-loss refractive index. Taking into account the limited set of analytical solutions  in optical waveguiding, the method presented here represents a very efficient resource of theoretical models on the matter. The approach is based on the finite-difference algorithm for higher-order supersymmetry introduced in~\cite{Mie00}, which summarizes the fundamental ingredients of supersymmetric quantum mechanics of any order~\cite{Mie04}. It is a generalization of the supersymmetric formulation introduced by Mielnik in 1984~\cite{Mie84}, which coincides with the Darboux transformation connecting two exactly solvable potentials in quantum mechanics. 

The organization of the paper is as follows. In Section~\ref{cap1} we briefly revisit the mathematical structure underlying the propagation of light in graded-index waveguides. The well-known identification of the Helmholtz equation with the Schr\"odinger eigenvalue problem is the point of departure to apply the supersymmetric finite-difference algorithm. In Section~\ref{uno} we provide general formulae to add guided TE modes one at a time in the dynamics of a given waveguide. Then, in Section~\ref{dos}, a new mechanism to produce complex-valued refractive indices with all-real eigenvalues is discussed. The achievement of this method is that the imaginary part of the new refractive index is such that its integration over the entire domain of definition is equal to zero, preserving in this form the total optical power of the system. Such property permits the acquisition of PT symmetry as a particular case. We also discuss briefly the bi-orthogonal structure that is necessary to properly normalize the guided TE modes of the new system. In Section~\ref{tres} we present immediate applications. We show that the method produces a wide variety of exactly solvable refractive indices in both the real- and complex-valued configurations. The former case permits the manipulation of the index profile to better-fit it with (possible) experimental observations. The second case leads to PT-symmetric as well as to non-PT-symmetric refractive indices, both configurations admitting all-real propagation constants in the point spectrum. Some conclusions are given in Section~\ref{conclu}. For the sake of completeness, we briefly revisit the generalities of the supersymmetric finite-difference algorithm in Appendix~\ref{ApA}.

\section{Mathematical Physics of Graded-Index Waveguides}
\label{cap1}

The Helmholtz equation
\be
\frac{\partial^2}{\partial x^2} \Phi_y (x,z) + \frac{\partial^2}{\partial z^2} \Phi_y (x,z) + k_0^2 n^2(x) \Phi_y(x,z)=0
\label{Hel1}
\ee
is useful to study TE modes propagating through  inhomogeneous dielectric materials with negligible magnetic polarizability and  dispersion. Here $k_0 (= w/c) =2\pi \lambda_0^{-1} $ stands for the wavenumber in vacuum, and the refractive index $n$ is assumed to depend only on the $x$-coordinate. Hence, $\Phi_y(x,z)$ represents an electric wave polarized in the $y$-direction, which propagates along the positive $z$-direction.

On the other hand, the paraxial Helmholtz equation
\be
\frac{1}{2k_0^2 n_*} \frac{\partial^2}{\partial x^2} E_y (x,z) + \frac{i}{k_0} \frac{\partial}{\partial z} E_y (x,z) +  [n(x)  - n_* ] E_y(x,z)=0
\label{tita}
\ee
results within the paraxial regime for weakly guiding media ($\vert n - n_* \vert \ll 1$), after using the ansatz $\Phi_y(x,z)= E_y (x,z)  \exp(i k_0 n_*  z)$. The number $n_* >0$ defines a reference refractive index. As $n(x)$ is independent of the $z$-coordinate, let us assume $E_y(x,z) = E(x) \exp(-i k_0 \varepsilon z)$, where $\varepsilon$ defines the spectrum propagation constants~\cite{Glo69}, then (\ref{tita}) is reduced to the eigenvalue problem
\be
\left[ -\frac{1}{2k_0^2 n_*} \frac{d^2}{dx^2} -n(x) +n_*  \right] E(x) = \varepsilon E(x), \quad \left. E(x) := E_y(x,z) \right\vert_{z=0}.
\label{Hel2}
\ee
Comparing Equation~(\ref{Hel2}) with the eigenvalue equation for one-dimensional stationary systems in quantum mechanics
\be
\left[ -\frac{\hbar^2}{2m} \frac{d^2}{dx^2} +V(x)  \right] \psi(x) = \mathcal{E} \psi(x),
\label{eigen}
\ee
we obtain the identification~\cite{Cru15a}
\be
k_0^2  \longleftrightarrow \frac{m w}{\hbar}, \quad \left[ -n(x) +n_* \right] n_* \longleftrightarrow  \frac{V(x)}{\hbar w}, \quad \varepsilon n_* \longleftrightarrow  \frac{ \mathcal{E}}{\hbar w}.
\label{iden}
\ee
Then, the TE modes $E_y(x,z) = E(x) \exp(-ik_0 \varepsilon z)$ may be associated with the solutions $\psi(x,t) = \psi(x) \exp(-i\mathcal{E} t/\hbar)$ of the related Schr\"odinger equation, where $z \longleftrightarrow t \sqrt{w \hbar/m}$.

The link between (\ref{Hel2}) and (\ref{eigen}) is even more clear after introducing the changes $x \rightarrow \chi /(k_0 \sqrt{2})$ and $x \rightarrow \chi \sqrt{\hbar/(2m w)}$, which gives
\be
\left[ - \frac{d^2}{d\chi^2} + (- n(\chi) + n_*) n_* \right] E(\chi) = \varepsilon n_* E(\chi), \quad \left[ - \frac{d^2}{d\chi^2}  + \mathcal{V} (\chi) \right] \psi(\chi) = \widetilde{\mathcal{E}} \psi(\chi),
\label{nueva1}
\ee
with $\hbar w \mathcal{V} = V$ and $\hbar w \widetilde{\mathcal{E}} = \mathcal{E}$.

In the sequel we take full advantage of the above mathematical relationship, providing solutions to the Helmholtz Equation (\ref{Hel2}) from the space state of the quantum mechanical problem (\ref{eigen}). Our aim is to design refractive indices $n(x)$ on demand, locating propagation constants at concrete positions of the point spectrum, and producing balanced gain-and-loss. Keeping this in mind we will apply the Darboux method~\cite{Rog02}, as it has been developed in supersymmetric quantum mechanics~\cite{Mie04}, to deform the eigenfunctions of a given refractive index into the eigenfunctions of another one. The approach is useful to add/eliminate a concrete number of eigenvalues to/from the point spectrum of $n(x)$, at the price of transforming $n(x)$ and $E(x)$ into new functions. Changing the point spectrum of $n(x)$ by only one eigenvalue $\varepsilon$ corresponds to the first-order Susy (Darboux) transformation. If the modification involves $k\geq 2$ eigenvalues, then one works with the $k$th-order  Susy transformation, which may be performed iterating $k$-times the Darboux transformation or deforming $n(x)$ in a single step (Darboux–Crum)~\cite{Mie04}. The Darboux method has been elegantly summarized in the finite-difference algorithm for higher-order supersymmetry introduced in~\cite{Mie00}, which is briefly revisited in Appendix~\ref{ApA} for completeness.

\subsection{Adding Propagation Constants under Prescription }
\label{uno}

We are interested in generating refractive indices $n(x)$ with exact solutions to the paraxial Helmholtz Equation (\ref{Hel2}). An elegant way to achieve this is offered by the finite-difference algorithm for higher-order supersymmetry~\cite{Mie00} revisited in Appendix~\ref{ApA}. The keystone is to have at hand at least one exactly solvable refractive index $n_0(x)$, the point spectrum of which is inherited to another refractive index $n_1(x; \epsilon)$, defined by the Darboux~transformation
\be
(-n_1(x; \epsilon) + n_{1,*}) \frac{n_{1,*}}{n_{0,*}} = -n_0(x) + n_{0,*} + \frac{\sqrt 2}{k_0 n_{0,*}} \beta'_1 (x; \epsilon),
\label{index1}
\ee
where $f'(x) =\frac{d}{dx} f(x)$, the superpotential $\beta_1(x; \epsilon)$ is solution of the nonlinear Riccati equation
\be
- \frac{1}{k_0 \sqrt 2} \beta'_1(x;\epsilon) + \beta_1^2(x; \epsilon) =  n_{0,*} \left[-n_0(x) +n_{0,*} - \epsilon \right],
\label{ricca}
\ee
and the factorization constant $\epsilon$ is to be determined (see Appendix~\ref{ApA} for details). 

To simplify notation, without loss of generality, hereafter we make $n_{k,*}=n_{0,*}:=n_*$, with $k=1,2,\ldots$ 

The refractive index (\ref{index1}) defines a new paraxial Helmholtz equation
\be
\left[ -\frac{1}{2k_0^2 n_*} \frac{d^2}{dx^2} -n_1(x; \epsilon) + n_* \right] E_{(1)} (x; \epsilon) = \varepsilon E_{(1)} (x; \epsilon),
\label{Hel2a}
\ee
the solutions of which are constructed from the eigenfunctions $E_{(0)}(x)$ of $n_0(x)$ as follows
\be
\mathcal{N}_{(1)}^{-1} E_{(1)}(x; \epsilon) = E'_{(0)} (x) + \beta_1(x,\epsilon) E_{(0)} (x),
\label{transform}
\ee
where $\mathcal{N}_{(1)}$ obeys normalization.

The point spectrum of $n_1(x)$ will be exactly the same as the one of $n_0(x)$, or it may include the factorization constant $\epsilon$ as an additional eigenvalue. In the latter case, the corresponding eigenfunction is written as follows
\be
E_{(1)}^M (x; \epsilon)   =  \mathcal{N}_{(1)}^M \exp \left[\int \beta_1(x; \epsilon) dx \right].
\label{missi}
\ee
The procedure described above can be iterated at will, see Appendix~\ref{ApA}. After $k$ steps one obtains a refractive index $n_k(x; \epsilon)$ that admits $k$ additional eigenvalues in its point spectrum with respect to the point spectrum of $n_0(x)$. 

Clearly, the profile of  $n_k(x;\epsilon)$ may be manipulated to obtain a concrete number of guided TE modes under prescription. For instance, suppose that $n_0(x)$ in (\ref{index1}) admits only $N$ guided modes, the finite-difference algorithm provides $n_1(x; \epsilon)$ with exactly the same propagation constants as $n_0(x)$ plus an additional one, located at $\epsilon$ in the point spectrum of $n_1(x; \epsilon)$ but absent in the spectrum of $n_0(x)$. 

Depending on the prescription, the propagation constants of the guided modes may be added one at a time, iterating the finite-difference algorithm as necessary, or using a single transformation if just one additional propagation constant is required. Conventional supersymmetric approaches include the new eigenvalue below the lowest eigenvalue of the previous spectrum. The latter obeys the fact that the oscillation theorem prohibits constructing regular superpotentials $\beta_k$ if the factorization constant $\epsilon$ is above the lowest propagation constant of $n_{k-1}$. Thus, viable factorization constants $\epsilon$ are at most equal to the lowest propagation constant of $n_{k-1}$. In contraposition, a mechanism producing complex-valued refractive indices as Darboux-deformed versions of real-valued ones can be managed by following the method introduced in~\cite{Ros15}. The novelty is that the eigenvalues of the new refractive indices will be all-real. Moreover, the additional eigenvalue $\epsilon$ can be incorporated at any position of the spectrum~\cite{Bla19}. The formulae (\ref{index1})--(\ref{missi}) still apply for complex-valued refractive indices, and are particularly important to generate balanced gain-and-loss.

\subsection{Balanced Gain-and-Loss Waveguides}
\label{dos}

Conventional supersymmetric approaches assume that the solution of the Riccati Equation (\ref{ricca}) is real-valued. However, complex-valued solutions are feasible even for real-valued refractive indices $n_0(x)$ and real factorization energies $\epsilon$. Indeed, the real and imaginary parts of Equation~(\ref{ricca}) give rise to the nonlinear differential equation
\begin{equation}
\left[-\frac{1}{2k_0^2n_*} \frac{d^2}{dx^2} -n_{0} (x) + n_* \right] \alpha(x; \epsilon) = \epsilon \alpha(x; \epsilon) - \frac{1}{2k_0^2 n_*} \frac{\lambda^{2}}{\alpha^{3}(x; \epsilon)}, \quad \lambda \in \mathbb R,
\label{FM6}
\end{equation}
which is named after Ermakov~\cite{Erm80}. Please note that (\ref{FM6}) coincides with the paraxial Helmholtz Equation (\ref{Hel2}) when $\lambda =0$. Then, the solutions may be expressed as the nonlinear superposition~\cite{Ros15}
\begin{equation}
\alpha(x;\epsilon)=\left[ a u_{(1); 1}^{2}(x; \epsilon)+b u_{(1); 1}(x; \epsilon) u_{(1); 2}(x;\epsilon)+c u_{(1); 2}^{2}(x; \epsilon) \right]^{1/2}, 
\label{FM7}
\end{equation}
where $u_{(1);1}$ and $u_{(1);2}$ are two-linearly independent solutions of (\ref{Hel2}) for $\varepsilon = \epsilon$. The $\alpha$-function is real-valued and free of zeros in $\mbox{Dom} \, n_0$ if the set $\{a,b,c \}$ is composited by positive numbers fulfilling
\begin{equation}
b^{2}-4ac=-4\lambda^{2}/W_{0}^{2},
\label{FM81}
\end{equation} 
where $W_0=W(u_{(1);1},u_{(1);2})= \operatorname{const}$ is the Wronskian of $u_{(1);1}$ and $u_{(1);2}$. 

Using (\ref{FM7}), the complex-valued superpotential acquires a simple form 
\be
\beta_1(x; \epsilon) = - \frac12 \frac{d}{dx} \ln v(x; \epsilon)  + i \frac{\lambda}{v(x; \epsilon)} = -\left[ \frac{v'(x; \epsilon) -i 2 \lambda}{2 v(x; \epsilon) } \right],
\label{mibeta}
\ee
where
\be
v(x;\epsilon)= a u_{(1); 1}^{2}(x; \epsilon)+b u_{(1); 1}(x; \epsilon) u_{(1); 2}(x;\epsilon)+c u_{(1); 2}^{2}(x; \epsilon).
\label{v}
\ee

Then, the Darboux transformation (\ref{index1}) gives the complex-valued refractive index 
\be
n_1(x; \epsilon) = n_0(x) + \frac{\sqrt{2}}{k_0 n_*} \frac{d}{dx} \left[ \frac{v'(x; \epsilon) -i 2 \lambda}{2 v(x; \epsilon) } \right].
\label{index3}
\ee
On the other hand, it may be shown that the imaginary part of $n_1$ satisfies the condition of {\em zero total area}~\cite{Jai17}:
\begin{equation}
\int_{\mathbb R} \mbox{Im}\,  n_1(x; \epsilon) dx = \left. - \frac{{\sqrt 2} \lambda}{k_0 n_*}
\frac{1}{ v(x; \epsilon) } \right \vert_{-\infty}^{+\infty} =0,
\label{zero}
\end{equation}
so the total optical power is conserved. Equation (\ref{zero}) implies a balanced interplay between gain and loss that does not depend on any symmetry of either $\operatorname{Im}  n_1(x;\epsilon)$ or $\operatorname{Re}  n_1(x;\epsilon)$. 

In contraposition to conventional supersymmetric approaches, the factorization energy $\epsilon$ can be positioned at any place in the point spectrum of $n_1$~\cite{Bla19}. Moreover, the missing state~(\ref{missi}), now written as
\be
E_{(1)}^M (x; \epsilon) = \frac{\mathcal{N}_{(1)}^M}{\sqrt{v (x; \epsilon) }} \exp \left[ i \lambda \int v^{-1}(x; \epsilon) dx \right],
\label{missi2}
\ee
is such that its real and imaginary parts are even and odd functions of the position-variable $x$, respectively.

We would like to emphasize that the nonlinear superposition (\ref{v}) marks a distance with conventional supersymmetric approaches. Indeed, we have already shown~\cite{Zel20d} that the superpotential (\ref{mibeta}) can be also written in the conventional logarithmic form $\beta(x; \epsilon) = -\frac{d}{dx} \ln w(x; \epsilon)$, where the coefficients of the linear superposition
\be
w(x; \epsilon) = a u_{(1),1} (x, \epsilon)  + \left( \frac{b}{2} -i \frac{\lambda}{W_0} \right) u_{(1),2} (x, \epsilon)
\label{w}
\ee
are ruled by the constraint (\ref{FM81}), with $a$ and $b$ complex numbers in general. Clearly, such a concrete combination of coefficients permits $n_1$ to satisfy the condition of zero total \mbox{area (\ref{zero})}, which defines it as a balanced gain-and-loss refractive index. 

\subsubsection{Bi-Orthogonality} 
\label{bisection}

The solutions of the paraxial Helmholtz Equation~(\ref{Hel2a}), with $n_1(x; \epsilon)$ given in (\ref{index3}), are obtainable from (\ref{transform}), (\ref{mibeta}) and (\ref{missi2}). However, while $E_{(1)}^M$ and all the TE modes $E_{(1)}$ are normalizable, they form a peculiar set since $E_{(1)}^M$ is orthogonal to all the $E_{(1)}$ but these last are not mutually orthogonal~\cite{Ros15} (such property is not a problem in the Hermitian case since all the new functions satisfy the conventional oscillation theorems). Nevertheless, the eigenfunctions of $n_1$ satisfy some properties of interlacing of zeros that permit the study of the related systems as if they were Hermitian~\cite{Jai17}. In this context, the bi-orthogonal set formed by the eigenstates $E_{(1)}$ of $n_1$, together with the eigenstates $\widetilde E_{(1)}$ of the complex-conjugated refractive index, written $n_1^C$, provide an extended space of states where all the basis elements are bi-orthonormal~\cite{Ros15,Ros18,Zel20a}. Indeed, the bi-product
\be
\left( \widetilde E_{(1); m}, E_{(1); n} \right) = \int_{\mathbb R} \widetilde E^C_{(1); m}
(x;\epsilon)  E_{(1); n} (x; \epsilon) dx 
\ee
is equal to zero if $n \neq m$, and serves to define the bi-norm $\vert \vert E_{(1);n} \vert \vert_B= \vert \vert \widetilde E_{(1);n} \vert \vert_B$ if $n=m$~\cite{Ros15}. With two possible normalizations at our disposal, $E_{(1)}/\vert \vert E_{(1)} \vert \vert$ and $E_{(1)}/\vert \vert E_{(1)} \vert \vert_B$, it is important to emphasize that the real and imaginary parts of the modes $E_{(1)}$ behave qualitatively equal in both normalizations~\cite{Zel20a}, although their bi-normalized values are usually larger than those obtained with the conventional normalization. Nevertheless, the differences become negligible as the excitation of the TE mode increases, see~\cite{Zel20a} for details. Note also that the notions of bi-product and  bi-norm introduced above coincide with the conventional definitions if $\lambda=0$.

\subsubsection{PT-Symmetric Case} 

The expression (\ref{index3}) represents a wide family of  balanced gain-and-loss refractive indices. A very interesting subset of such family is integrated by the so-called parity-time (PT) symmetric refractive indices. Recalling that invariance under parity and time‐reversal transformations requires $n(x) = n^C(-x)$ in quantum mechanics~\cite{Ben98}, we realize that the initial refractive index $n_0(x)$ should be parity-invariant $n_0(x) = n_0(-x)$ to facilitate the construction of PT-symmetric refractive indices $n_1(x; \epsilon)$. On the other hand, assuming real-valued transformation functions $u_{(1);1}$ and $u_{(1); 2}$, we may take $b=0$ in (\ref{v}) to obtain the quadratic form
\be
v_{PT}(x;\epsilon)= a u_{(1); 1}^{2}(x; \epsilon) + c u_{(1); 2}^{2}(x; \epsilon).
\label{v2}
\ee
The straightforward calculation shows that using this function in (\ref{index3}) one obtains a complex-valued graded index that is PT-symmetric. 

Please note that $b=0$ implies $ac = \lambda^2/W_0^2$ in (\ref{FM81}). To simplify notation, without loss of generality, let us make $a=c=\lambda/\vert W_0 \vert$ in (\ref{v2}), where $\vert W_0 \vert$ stands for the modulus of $W_0$. Then (\ref{index3}) yields
\begin{multline}
n_{PT}(x; \epsilon) = n_0(x) + \frac{1}{ {\sqrt 2} k_0 n_*}  \frac{d^2}{dx^2} \ln \left[ u_{(1); 1}^{2}(x; \epsilon) +  u_{(1); 2}^{2}(x; \epsilon) \right]\\
 - i  \frac{{\sqrt 2} \vert W_0 \vert}{k_0 n_*} \frac{d}{dx}\left[ u_{(1); 1}^{2}(x; \epsilon) +  u_{(1); 2}^{2}(x; \epsilon) \right]^{-1}.
\label{index4}
\end{multline}
Observe that the $w$-configuration (\ref{w}) leads to the same result provided $a=-i\lambda/W_0^C$.

\subsubsection{Recovering the Real-Valued Case} 
\label{twosuper}

As indicated above, if $\lambda=0$ the superpotential (\ref{mibeta}) is reduced to its Hermitian configuration, which produces real-valued indices only. Revisiting the constraint (\ref{FM81}) we see that $\lambda=0$ implies $b^2 = 4ac$, and thus $b = \pm 2 \sqrt{ac}$. We obtain the linear superpositions $\alpha_{\pm} = \sqrt{a} u_{(1);1} \pm \sqrt{c} u_{(1); 2}$, so we arrive at the conventional superpotentials
\be
\beta_R(x; \epsilon; \pm)= -\frac{d}{dx} \ln \left[\sqrt{a} u_{(1);1} (x; \epsilon) \pm \sqrt{c} u_{(1); 2} (x; \epsilon) \right],
\label{betar}
\ee
where $a$ and $c$ are such that $\beta$ is free of singularities in $\operatorname{Dom}n_0$. Therefore, from (\ref{index1}) one has the two-parametric family of real-valued graded indices
\be
n_R(x; \epsilon; \pm) = n_0(x) +\frac{\sqrt 2}{k_0 n_*} \frac{d^2}{dx^2} \ln \left[\sqrt{a} u_{(1);1} (x; \epsilon) \pm \sqrt{c} u_{(1); 2} (x; \epsilon) \right].
\label{index2}
\ee


\section{Applications}
\label{tres}

The method developed in previous sections may be applied to practically any exactly solvable refractive index $n_0(x)$. The expression $n_k(x; \epsilon)$, obtained at the $k$th step, actually represents a very wide family of new exactly solvable refractive indices with $k$ additional propagation constants in their point spectrum with respect to $n_0(x)$. Even more important is the fact that $n_k(x; \epsilon)$ can be constructed to be a real- or complex-valued function. In any case, the propagation constants belonging to the point spectrum of $n_k(x; \epsilon)$ are all-real. Next, we provide a very useful example of the applicability of our approach. We use the mathematical solutions associated with $n_0=0$ to produce diverse families of cosh-like refractive indices admitting the presence of a given number of guided TE modes. The results include complex-valued refractive indices that are not limited to the parity-invariant case.

From now on, for the sake of simplicity, the expressions of the refractive index profiles $n_0(x)$ and $n_k(x; \epsilon)$ are considered up to the additive constant $n_*$.

\subsection{Adding Guided Modes One at a Time}

\begin{figure}[htb]
\centering
\subfloat[][$n_R(x; \kappa)$]{\includegraphics[width=0.3\textwidth]{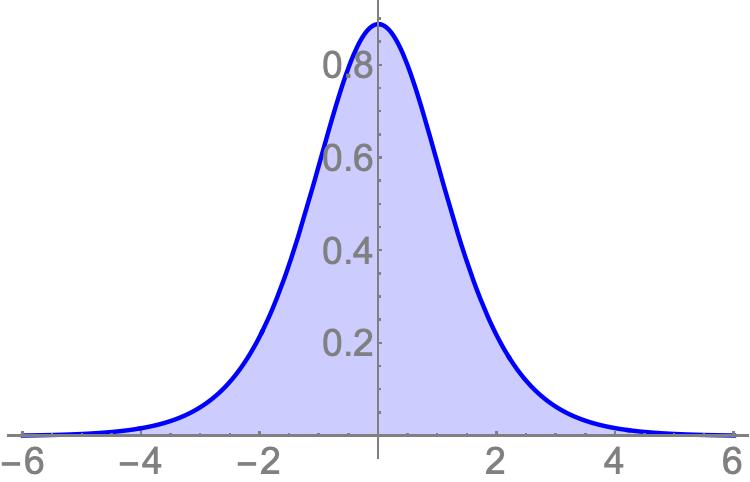}}
\hskip4ex
\subfloat[][$n_{PT}(x;\kappa)$]{\includegraphics[width=0.3\textwidth]{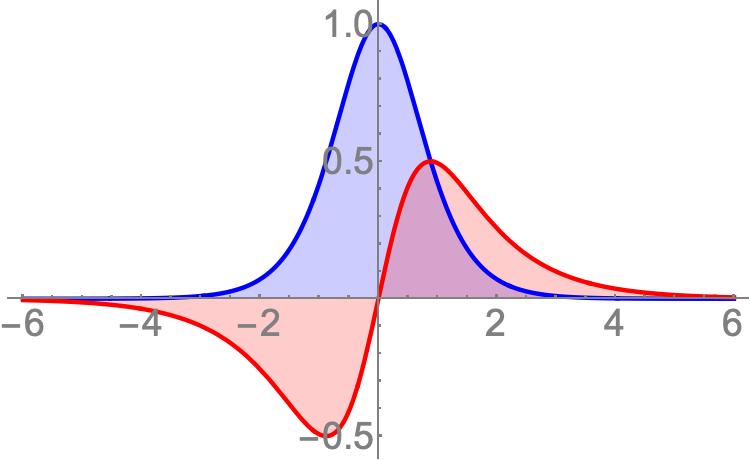}}

\caption{\footnotesize Cosh-like refractive index (\ref{betar2}) for $\kappa=2/3$ 
(\textbf{a}). Parity-time symmetric refractive index (\ref{ptindex}) for $\kappa=1/2$ (\textbf{b}). In both cases $x_0=0$, $k_0=1/\sqrt{2n_*}$, and the horizontal axis is mounted on $n_*$. These graded refractive indices admit the presence of only one guide TE mode of propagation constant $\varepsilon=-4/9$ and $\varepsilon =-1/4$, respectively. In both cases the real part is in blue while the imaginary part is in red.
}
\label{A}
\end{figure}

The fundamental solutions of the paraxial Helmholtz equation for $n_0=0$ are well known. For positive factorization constants \mbox{$\epsilon = k^2>0$} we write $u_1=e^{ik(x-x_0)}$ and \mbox{$u_2=e^{-ik(x-x_0)}$}, with $W_0=-i2k$. However, the above expressions yield sinusoidal refractive indices $n_1$~\cite{Ros15}, which are out of the scope of the present work. Here we make $k=i \kappa$ to obtain negative factorization constants \mbox{$\epsilon = -\kappa^2$},~therefore 
\be
v(x; \kappa)= a e^{2\kappa (x-x_0)} + c e^{-2 \kappa (x-x_0)} + b, \quad b^2-4ac = - \lambda^2/\kappa^2.
\label{uno1}
\ee
To simplify notation let us make $a=c$. Then $v(x; \kappa) = 2a \cosh[2 \kappa (x-x_0)] +b$, with $b^2=4 a^2-\lambda^2/\kappa^2$. The superpotential (\ref{mibeta}) acquires the form
\be
\beta_1(x; \kappa) = -\left[ \frac{ \kappa \sinh[2\kappa (x-x_0)] -i \frac{\lambda}{2a} }{ \cosh[2 \kappa (x-x_0)] + \frac{b}{2a} } \right], \quad b^2=4 a^2-\frac{\lambda^2}{\kappa^2},
\label{mibeta2}
\ee
so the refractive index (\ref{index3}) is in this case
\be
n_1(x; \kappa) = \frac{ (2\kappa)^2 \left( 1 + \frac{b}{2a} \cosh[2\kappa (x-x_0)] \right) + i \frac{\lambda}{a} (2\kappa) \sinh[2 \kappa (x-x_0)]}{{\sqrt 2} k_0 n_* \left( \cosh [2 \kappa (x-x_0)] + \frac{b}{2a} \right)^2 }.
\label{index}
\ee
The complex-valued graded refractive index (\ref{index}) allows the presence of only one guided TE mode, obtained from (\ref{missi}) in the form
\be
E_{(1)}^M (x; \kappa)   = \frac{ \mathcal{N}_{(1)}^M }{\sqrt{\cosh [2 \kappa (x-x_0)] + \frac{b}{2a} } } \exp \left\{ -\tfrac{i}{4a} \arctan \left( \left( \tfrac{b}{2a}-1 \right) \tanh [\kappa (x-x_0)] \right) \right\}.
\label{mimissi}
\ee
Following the indications of the previous section, let us make $\lambda=0$ and $b=2a$ in (\ref{mibeta2}) and (\ref{index}) to obtain 
\be
\beta_R(x; \kappa) = - \kappa \tanh [\kappa (x-x_0)], \quad n_R (x;\kappa) = \frac{2 \kappa^2}{{\sqrt 2} k_0 n_* \cosh^2 [\kappa (x-x_0) ]},
\label{betar2}
\ee
which are the well-known expressions for the cosh-like refractive index. The function $n_R(x;\kappa)$ is depicted in Figure~\ref{A}a for a representative propagation constant $\varepsilon$, which may be located at any position $\varepsilon_1 = -\kappa_1^2$ since it is the only one eigenvalue in the point spectrum.

For $b=0$ and $a= \frac{\lambda}{2\kappa}$ the formulae (\ref{mibeta2}) and (\ref{index}) give rise to the PT-symmetric expressions
\be
\beta_{PT}(x; \kappa) = -\left( \tfrac{\kappa \sinh [2 \kappa (x-x_0)] - i\kappa}{\cosh [2 \kappa (x-x_0)] } \right),  \quad
 n_{PT} (x; \kappa) = \tfrac{(2\kappa)^2  \left( 1+ i \sinh [2\kappa (x-x_0)] \right)}{{\sqrt 2} k_0  n_* \cosh^2 [2\kappa (x-x_0) ]}.
\label{ptindex}
\ee
The PT-symmetric refractive index (\ref{ptindex}) is shown in Figure~\ref{A}b for a representative propagation constant $\varepsilon$. As in the previous case, this eigenvalue can be positioned at will in the negative part of the real axis.

If we repeat the procedure, assuming now that $n_1(x; \kappa_1)$ has been already fixed, the finite-difference algorithm provides an immediate superpotential (see details in \mbox{Appendix~\ref{ApA}}):
\be
\beta_2(x; \kappa_1, \kappa) = - \beta_1(x; \kappa_1) + \frac{{\sqrt 2} k_0 n_*(\kappa_1^2 -\kappa^2)}{ \beta_1 (x; \kappa_1) - \beta_1(x; \kappa)},
\label{mibeta3}
\ee
where $\kappa_1$ and $\beta_1(x; \kappa_1)$ have been fixed in the previous step. Deciding the concrete value of $\kappa$, as well as the analytical form of $\beta_1(x; \kappa)$ in (\ref{mibeta2}), the above equation provides the new refractive index 
\be
n_2(x; \kappa_1,\kappa) = n_1(x; \kappa_1) -\frac{\sqrt 2}{k_0 n_*} \beta'_2(x; \kappa_1, \kappa)
= -\frac{d}{dx} \left[\frac{2 (\kappa_1^2 -\kappa^2)}{ \beta_1(x; \kappa_1) - \beta_1(x; \kappa)} \right],
\label{supern}
\ee
where we have used (\ref{index1}) with $n_0(x)=0$. 

\begin{figure}[h]
\centering
\subfloat[][$x_0=x_1=0$]{\includegraphics[width=0.3\textwidth]{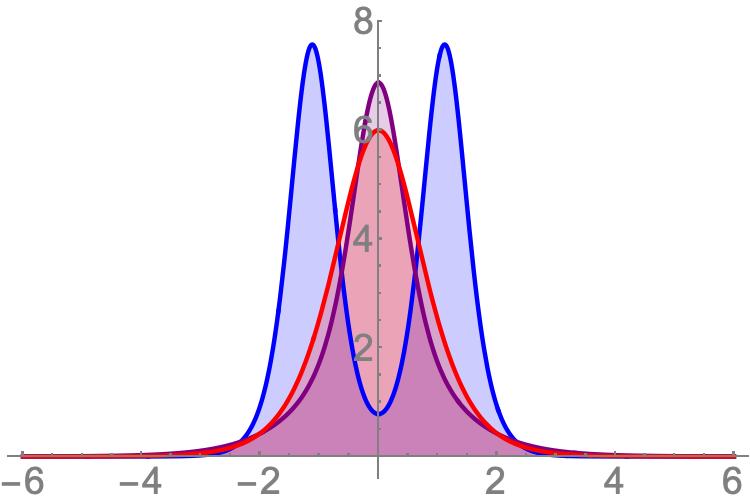}}
\hskip4ex
\subfloat[][$x_0\neq 0$, $x_1\neq 0$]{\includegraphics[width=0.3\textwidth]{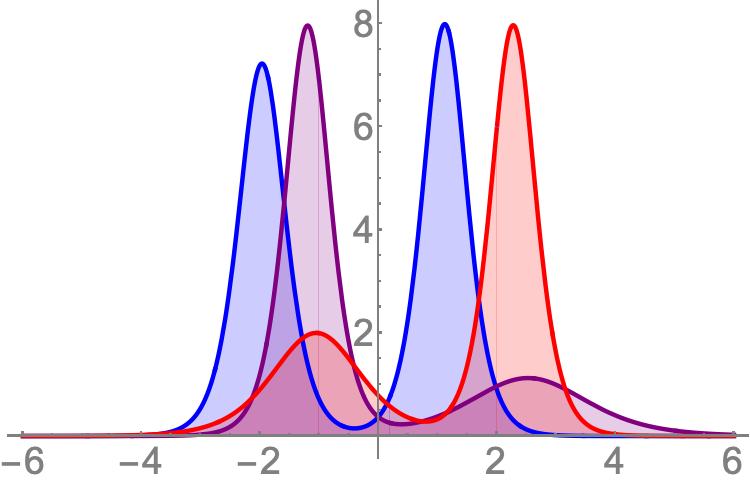}}

\caption{\footnotesize Real-valued refractive index $n_R(x;\kappa_1,\kappa)$ with symmetric (\textbf{a}) and non-symmetric (\textbf{b}) profile, see Equation~(\ref{mireal}). The point spectrum is composited by only two propagation constants $\{ \varepsilon, \varepsilon_1\}$, explicitly $\{-(1.9)^2,-4 \}$, $\{ -\frac9{16},-4\}$, and $\{-1,-4\}$ for curves in blue, purple and red, respectively. In (\textbf{b}) the points $(x_0,x_1)$ are $(0.2,-1)$, $(-1,2)$, and $(2,-0.5)$, following the color code indicated above.
}
\label{B}
\end{figure}

At the present stage, we have incorporated two propagation constants, so the point spectrum of $n_2(x; \kappa_1,\kappa)$ is composited by the eigenvalues $\varepsilon_1=-\kappa_1^2$ and $\varepsilon =-  \kappa^2$. However, some caution is necessary if the first step was addressed to produce $n_R(x;\kappa_1)$ and we are looking for a second real-valued refractive index $n_R(x; \kappa_1, \kappa)$. In such a case the inequality $\varepsilon < \varepsilon_1$ must be satisfied to obtain regular functions $n_R(x; \kappa_1, \kappa)$. Moreover, in such case, it may be shown~\cite{Mie00} that it is better to combine the two different real-valued superpotentials $\beta_R(x; \kappa; \pm)$. The case ``$+$'' is reported in Equation~(\ref{betar2}), the case ``$-$'' corresponds to the complementary expression $\beta_R(x; \kappa; -) = -\kappa \coth[\kappa (x-x_0)]$, see Section~\ref{twosuper}. We therefore arrive at the real-valued graded index 
\be
n_R(x;\kappa_1,\kappa) = \frac{2(\kappa_1^2 -\kappa^2) \left( \kappa_1^2 \operatorname{csch}^2 [ \kappa_1 (x-x_1)] + \kappa^2 \operatorname{sech}^2 [\kappa (x-x_0)] \right) }{\left( -\kappa_1 \coth [\kappa_1 (x-x_1)] + \kappa \tanh [\kappa (x-x_0)] \right)^2}.
\label{mireal}
\ee
The behavior of $n_R(x;\kappa_1,\kappa)$ is shown in Figure~\ref{B} for different spectra $\{\varepsilon, \varepsilon_1\}$ and constants $x_0$ and $x_1$. 

\begin{figure}[h]
\centering
\subfloat[][$x_0=x_1=0$]{\includegraphics[width=0.3\textwidth]{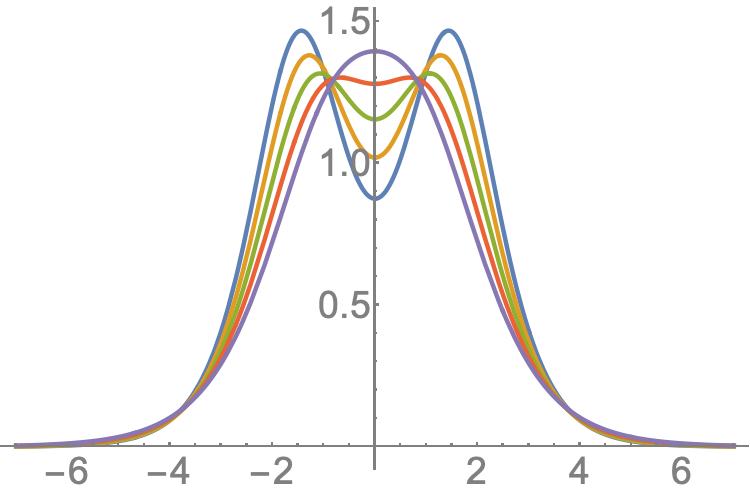}}
\hskip3ex
\subfloat[][$x_0=0, x_1 \neq 0$]{\includegraphics[width=0.3\textwidth]{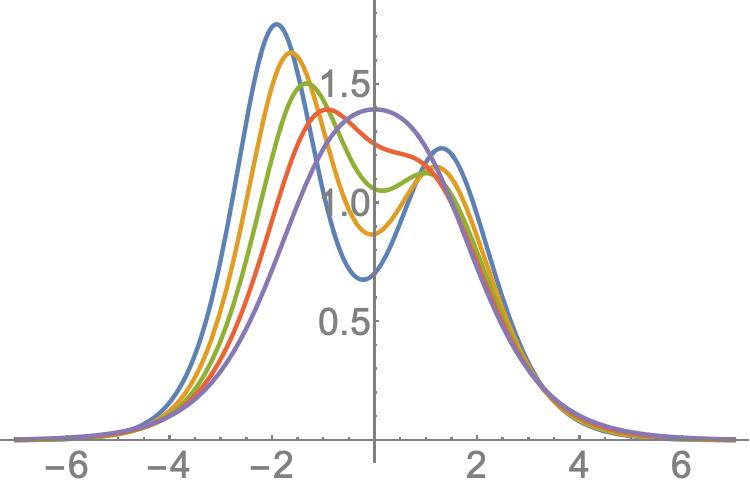}}
\hskip3ex
\subfloat[][$\{\kappa_1, \kappa\}$; $x_0=0, x_1 \neq 0$]{\includegraphics[width=0.3\textwidth]{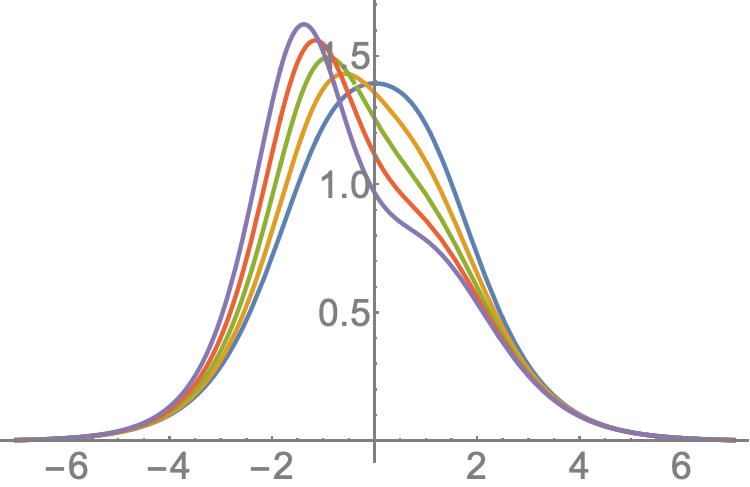}}

\caption{\footnotesize The real-valued refractive index $n_R(x; \kappa_1,\kappa)$ may be produced with a dent if $\kappa = \kappa_1- \varrho$, with $0 \leq \varrho \ll 1$, in Equation~(\ref{mireal}). This may be symmetrical (\textbf{a}) or asymmetrical (\textbf{b}). Both cases correspond to the point-spectra $\{-\kappa_1^2, -\kappa^2 \}$, where $\kappa_1=1$ and $k=0.55,0.6,0.65,0.7,0.75$, top to bottom curves as they are viewed at $x=0$, respectively. In (\textbf{b}), the same curves are evaluated with $x_1=0, -0.2,-0.4,-0.6$, and $-0.8$. In (\textbf{c}) the spectrum is fixed, with $\kappa_1=1$ and $\kappa=0.55$. The displacement $x_1$ takes the values indicated in (\textbf{b}). The dents in (\textbf{a},\textbf{b}), as well as the deformations (\textbf{c}), are deliberately produced in the manufacture of actual refractive indices, see for instance~\cite{Oko82}. }
\label{dent}
\end{figure}

Remarkably, $\varepsilon$ and $\varepsilon_1$ characterize the global profile of the function (\ref{mireal}). Indeed, for $\kappa_1 \gg \kappa$ and $x_0=x_1=0$, the refractive index $n_R(x;\kappa_1,\kappa)$ acquires a bell-shaped form. However, a valley arises at the top of such curve if $\kappa = \kappa_1- \varrho$, with $0 \leq \varrho \ll 1$. The dent is more pronounced as $\varrho \rightarrow 0$, separating the initial bell-like curve into a pair of bell-shaped ones. At the very limit, the new curves have moved in opposite directions toward the domain edges $\pm \infty$. Quite interestingly, actual waveguides are manufactured by including such dent, ``sometimes for reducing the internal mechanical stress due to the gradient of dopant concentration, and sometimes for reducing the multimode dispersion''~\cite{Oko82}, p.~83. With this in mind, Figure~\ref{dent} shows the exploration of the parameters that characterize $n_R(x;\kappa_1,\kappa)$, addressed to produce different dent configurations in the refractive index. These may be completely symmetrical as in Figure~\ref{dent}a or asymmetrical, as shown in  Figure~\ref{dent}b. For $\kappa_1 > \kappa$, local deformations may be produced by tuning the displacement parameters $x_0$ and $x_1$, see Figure~\ref{dent}c. 

The ordering problem suffered by the propagation constants in the construction of $n_R(x;\kappa_1,\kappa)$ is easily avoided by considering any superpotential (\ref{mibeta2}) with $\lambda \neq 0$ in either of the two steps. For instance, as in the previous example, assume that $n_R(x;\kappa_1)$ has been fixed in the first step. To include the second eigenvalue $\varepsilon$ this time we use the complex-valued superpotential $\beta_{PT}(x; \kappa)$ introduced in (\ref{ptindex}). The new refractive index (\ref{supern}) is now complex-valued, given by
\be
n_2(x; \kappa_1,\kappa) = \frac{2(\kappa_1^2 -\kappa^2) f(x; \kappa_1,\kappa) }
{ g^2 (x; \kappa_1,\kappa) },
\label{gral1}
\ee
where 
\be
f(x; \kappa_1,\kappa) = -\kappa_1^2 \operatorname{sech}^2(\kappa_1 x) + 2 \kappa^2 \operatorname{sech}^2(2 \kappa x) + i 2 \kappa^2 \tanh (2 \kappa x) \operatorname{sech}(2 \kappa x)
\label{gral2}
\ee
and
\be
g(x; \kappa_1,\kappa) = - \kappa_1 \tanh(\kappa_1 x) + \kappa \tanh (2 \kappa x) - i \kappa \operatorname{sech}(2\kappa x).
\label{gral3}
\ee
In (\ref{gral2}) and (\ref{gral3}) we have omitted the displacement constants $x_0$ and $x_1$ for the sake of simplicity.

\begin{figure}[htb]
\centering
\subfloat[][$\varepsilon_1 \Rightarrow \varepsilon$]{\includegraphics[width=0.3\textwidth]{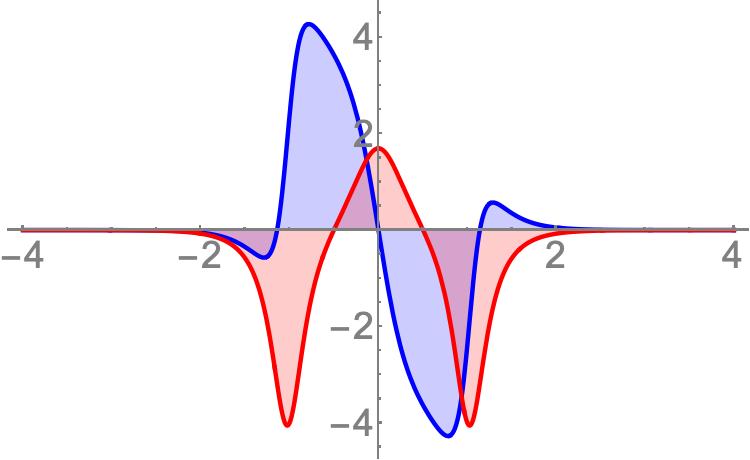}}
\hskip4ex
\subfloat[][$\varepsilon_1 \Leftarrow \varepsilon$]{\includegraphics[width=0.3\textwidth]{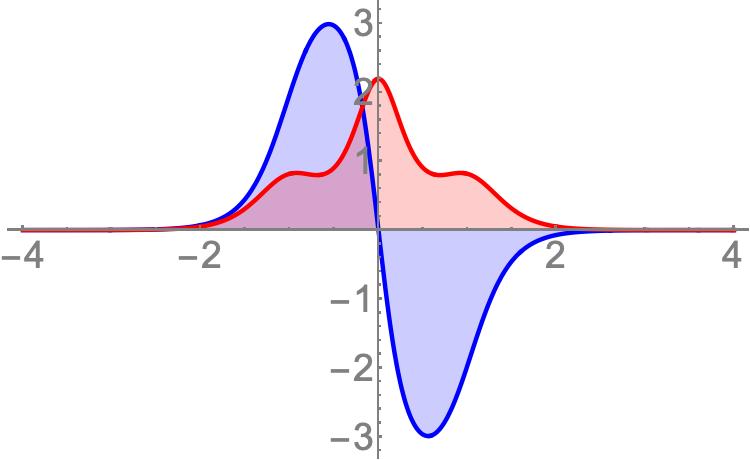}}

\caption{\footnotesize PT-symmetric version of the complex-valued refractive index $n_1(x; \kappa_1, \kappa)$ introduced in Equation~(\ref{gral1}). In contrast with the real-valued case $n_R(x; \kappa_1, \kappa)$, the propagation constants can be added in arbitrary order to the point spectrum $\{ \varepsilon, \varepsilon_1\}$. Nevertheless, although the PT symmetry is preserved, the profile of $n_1(x; \kappa_1, \kappa)$ is affected by the change $\varepsilon_1 \leftrightarrow \varepsilon$. The point spectrum is $\{-(1.9)^2,-4\}$. In both cases $x_0=x_1=0$, with the real and imaginary parts in blue and red, respectively. 
}
\label{C}
\end{figure}

\begin{figure}[htb]
\centering
\subfloat[][$\varepsilon_1 \Rightarrow \varepsilon$]{\includegraphics[width=0.3\textwidth]{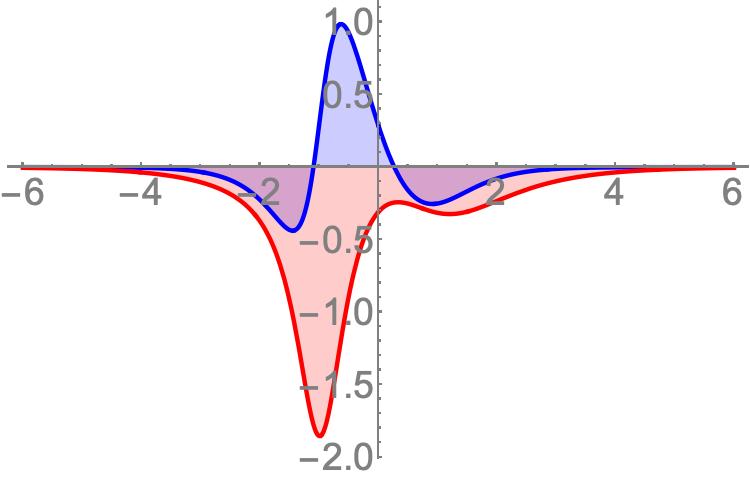}}
\hskip4ex
\subfloat[][$\varepsilon_1 \Leftarrow \varepsilon$]{\includegraphics[width=0.3\textwidth]{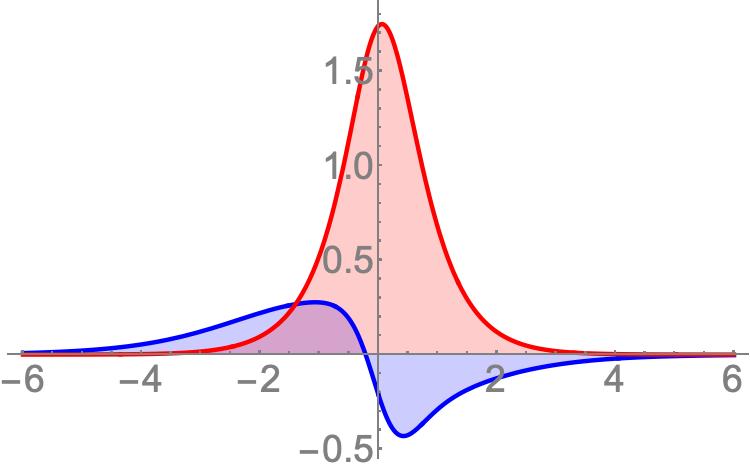}}

\caption{\footnotesize Same as in Figure~\ref{C}, with $x_0= -0.5$ and $x_1=0$. In this case, the complex-valued refractive index $n_1(x; \kappa_1, \kappa)$ is not invariant under PT-transformations.
}
\label{D}
\end{figure}

As $n_R(x; \kappa_1)$ is parity-invariant $n_R(x; \kappa_1) = n_R(-x; \kappa_1)$, the parameters of $n_2(x; \kappa_1,\kappa)$ in (\ref{gral1}) can be managed to obtain a PT-symmetric refractive index $n_{PT} (x; \kappa_1,\kappa)$. The result is shown in Figure~\ref{C}a for the process in which we add first $\varepsilon_1$ and then $\varepsilon$, with $\varepsilon_1 > \varepsilon$. The reversed process is shown in Figure~\ref{C}b. Please note that although the profile of $n_{PT} (x; \kappa_1,\kappa)$ changes, the PT symmetry is preserved under the change $\varepsilon_1 \leftrightarrow \varepsilon$. The same expression (\ref{gral1}) gives rise to refractive indices that are not invariant under parity-time transformations, as exhibited in Figure~\ref{D}.

We have already mentioned that the procedure may be repeated at will. At the $k$th step, the method provides a set of superpotentials $\beta_k$ that are available for the finite-difference algorithm, addressed to elaborate the step $k+1$. The case considered in this section takes the null function $n_0(x)=0$ as the initial refractive index. The propagation constants are added one at a time to arrive at the point spectrum $\{\varepsilon_1, \varepsilon_2, \ldots, \varepsilon_{k-1},\varepsilon \}$, which may be decided under prescription. The refractive indices constructed in this form admit $k$ guided TE modes, generated from the initial missing state $E^M_{(1)}(x; \varepsilon)$, via the rule (\ref{transform}). These modes obey the bi-product introduced in Section~\ref{bisection}, which also defines a proper bi-norm that coincides with the conventional norm if $\lambda=0$.

\subsection{Manipulating a Set of Guided Modes at Once}

To complete the revision of immediate applications, consider the cosh-like refractive~index
\be
n_0(x, m) = \frac{m(m+1) \kappa^2}{\cosh^2(\kappa x)}, \quad m=1,2,\ldots
\label{mipot}
\ee
Potentials $V(x)=-n_0(x, m)$ form the subset of transparent P\"oschl–Teller potentials in quantum mechanics. The solutions of the Schr\"odinger equation for the entire family are well known~\cite{Dia99,Flu74,Cru08}, including resonances and anti-bound states~\cite{Cev16,Civ19}. Our interest in $n_0(x, m)$ obeys the fact that this refractive index admits exactly $m$ guided TE modes, defined by the quadratic rule~\cite{Dia99,Flu74,Cru08}
\be
\varepsilon_{m,\ell}= -\kappa^2(m-\ell)^2, \quad \ell= 0,1,\ldots, m-1, \quad m=\operatorname{fixed}.
\label{spectrum}
\ee
The finite-difference algorithm will provide $k$ additional eigenvalues at the $k$th iteration, so the spectrum of $n_k(x, m;\kappa)$ is composited by two finite subsets $\{\varepsilon_i \} \cup \{\varepsilon_{m,\ell} \}$, with $i=1,2,\ldots,k$. As we have shown in the previous section, depending on the 1-step superpotentials $\beta_1(x,m; \varepsilon)$ and the factorization constants $\varepsilon$, the new eigenvalues $\varepsilon_i$ may be positioned at arbitrary places of the initial spectrum $\{\varepsilon_{m,\ell} \}$. 

The fundamental basis of solutions is in this case provided by the functions~\cite{Dia99,Flu74}
\be
u_1(x; \kappa) = ( \cosh \kappa x )^{m+1} \, {}_2F_1 (a , b, \tfrac12; -\sinh^2 \kappa x),
\label{sol1}
\ee
and
\be
u_2(x; \kappa) = (\sinh \kappa x )  ( \cosh \kappa x )^{m+1} \, {}_2F_1 (a +\tfrac12, b+\tfrac12, \tfrac32; -\sinh^2 \kappa x),
\label{sol2}
\ee
where
\be
a= \frac{m+1}{2} - \frac{\sqrt{ \vert \varepsilon \vert}}{2\kappa}, \quad b= \frac{m+1}{2} + \frac{\sqrt{ \vert \varepsilon \vert}}{2\kappa}.
\label{sol3}
\ee

To construct the complex-valued superpotential (\ref{mibeta}), a first function $v(x;\kappa)$ is easily achieved by noticing that the hypergeometric function ${}_2F_1$ is reduced to the identity if $a=0$. From (\ref{sol3}) we immediately realize that $\vert \varepsilon \vert = \kappa^2(m+1)^2$ produces such a result. Remarkably, the latter value is in correspondence with the spectral rule (\ref{spectrum}) if $\ell=-1$. Thus, we are in position of adding the eigenvalue $\varepsilon = \varepsilon_{m,-1}$ to the initial spectrum $\{\varepsilon_{m,\ell} \}$. The resulting refractive index $n_1(x,m;\kappa)$, obtained from Equation~(\ref{index3}), may be chosen to be either real, complex-valued or PT-symmetric.

\begin{figure}[htb]
\centering
\subfloat[][$\{\varepsilon_M,\varepsilon_{1,0}\}$]{\includegraphics[width=0.3\textwidth]{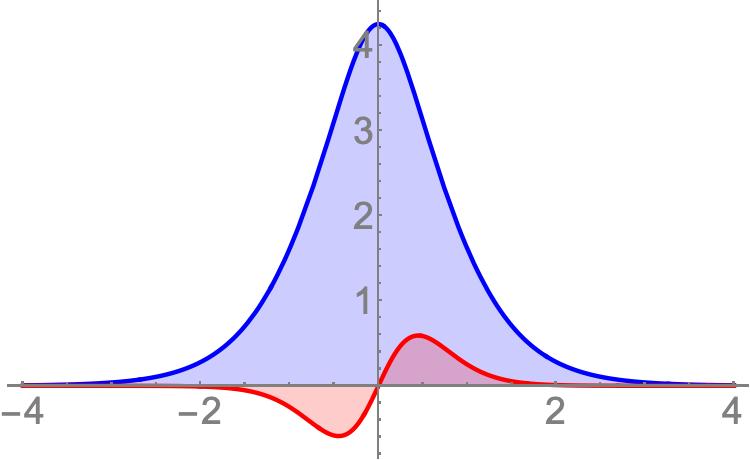}}
\hskip2ex
\subfloat[][$\{\varepsilon_M,\varepsilon_{2,0}, \varepsilon_{2,1}\}$]{\includegraphics[width=0.3\textwidth]{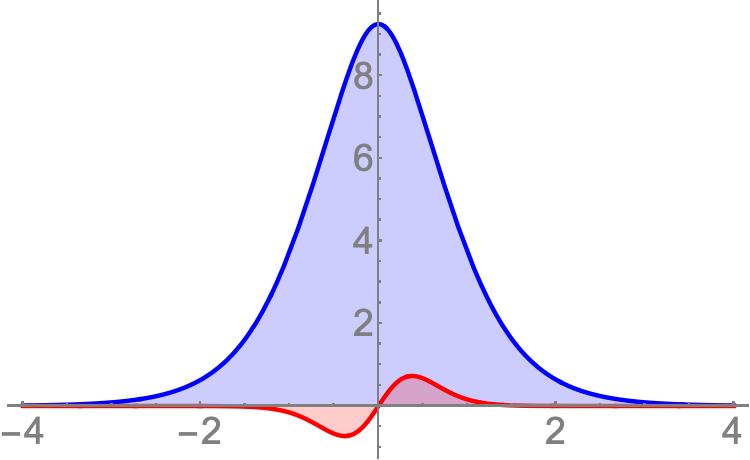}}
\hskip2ex
\subfloat[][$\{\varepsilon_M,\varepsilon_{3,0}, \varepsilon_{3,1},\varepsilon_{3,2}\}$]{\includegraphics[width=0.3\textwidth]{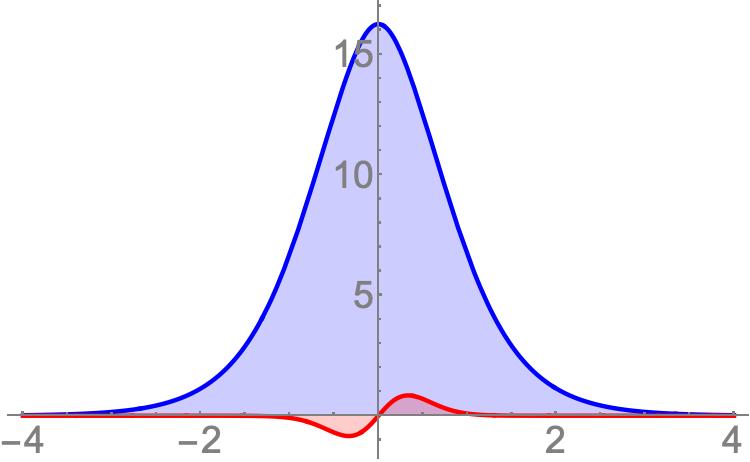}}

\caption{\footnotesize Complex-valued cosh-like refractive indices  $n_1(x,m;\kappa)$ exhibiting PT symmetry. The point spectrum is finite, including the eigenvalues indicated in captions. The spectral distribution is given by $\varepsilon_{m,\ell} = -\kappa^2(m-\ell)^2$, with $\ell=0,1,\ldots,m-1$, and $m \geq 1$ denoting the number of eigenvalues in the initial spectrum. In all cases $\varepsilon_M$ is located at $-\kappa^2 (m+1)^2$.
}
\label{E}
\end{figure}

\begin{figure}[htb]
\centering
\subfloat[][$\{\varepsilon_M,\varepsilon_0\}$]{\includegraphics[width=0.3\textwidth]{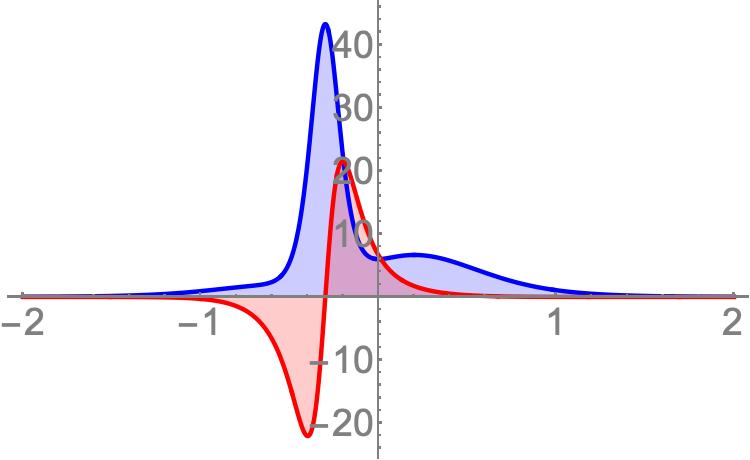}}
\hskip2ex
\subfloat[][$\{\varepsilon_M,\varepsilon_0, \varepsilon_1\}$]{\includegraphics[width=0.3\textwidth]{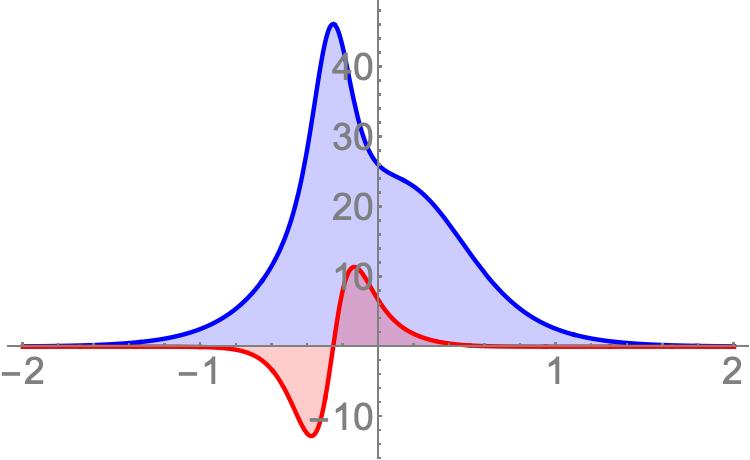}}
\hskip2ex
\subfloat[][$\{\varepsilon_M,\varepsilon_0, \varepsilon_1,\varepsilon_2\}$]{\includegraphics[width=0.3\textwidth]{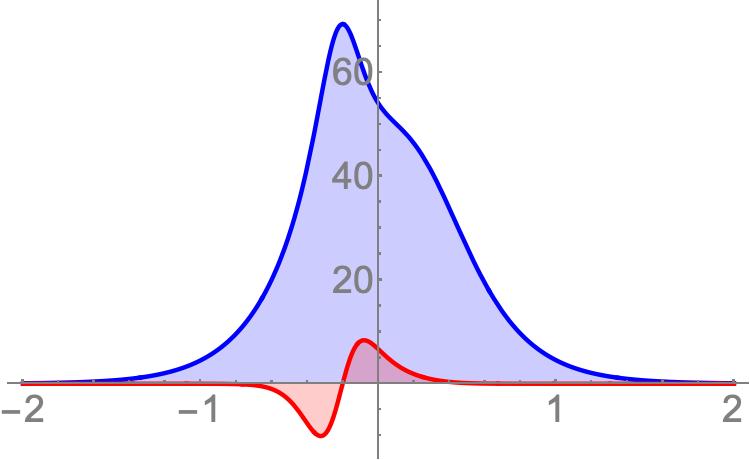}}

\caption{\footnotesize Same as in Figure~\ref{E}, with non-PT symmetry.
}
\label{F}
\end{figure}

In Figure~\ref{E} we have depicted the case in which $n_1(x,m;\kappa)$ exhibits PT symmetry.   In Figure~\ref{E}a we started with $n_0(x,1)$, which admits only one guided TE mode, the one associated with $\varepsilon_{1,0} $. The spectrum of the resulting refractive index $n_1(x,1;\kappa)$ is therefore integrated by $\varepsilon_M= -(2\kappa)^2$ and $\varepsilon_{1,0} =-\kappa^2$. Figure~\ref{E}b considers the initial spectrum $\varepsilon_{2,0} =-(2\kappa)^2$, $\varepsilon_{2,1} =-\kappa^2$, and includes the missing state $\varepsilon_M= -(3\kappa)^2$. Similarly for Figure~\ref{E}c. The configuration where the new refractive index is not PT-symmetric is shown in Figure~\ref{F}.

\section{Discussion of Results and Conclusions}
\label{conclu}

We have provided new exactly solvable models for optical waveguiding. Applying the supersymmetric finite-difference algorithm~\cite{Mie00}, we have generated a wide family of refractive indices whose point spectrum can be designed under prescription. The family includes refractive indices in both the real- and complex-valued configurations, the latter admitting all-real eigenvalues (propagation constants) in their point spectrum. We have shown that the spectral distribution may be organized in arbitrary form if it is constructed adding one at a time the eigenvalues such that complex-valued superpotentials are included. The result is relevant since such a property seems to be unnoticed in optical supersymmetry until the present work, although we have already reported this possibility in quantum mechanics~\cite{Bla19}. 

One of the main results presented in this work shows that the index profile strongly depends on the factorization constants that are incorporated. In particular, adding two of them, either in a single step or in a twice iterated movement, one may produce a dent over the top of the profile that is in complete agreement with actual manufacture of refractive indices~\cite{Oko82}, see Figure~\ref{dent}. The phenomenon is not exclusive of the real-valued indices so produced since it is also admissible in the complex-valued case for the real part of the PT-symmetric refractive indices, see Figure~\ref{B}. Considering that ``in the manufacture and evaluation of optical fibers, the measurement of the index profile is one of the most important steps''~\cite{Oko82}, our results may be useful to analyze the data obtained from such measurements.

Another of our results shows that refractive indices admitting a given number of guided TE modes, such as the sech-like ones, can be deformed to admit an additional guided mode, the propagation of which can be positioned anywhere in the point spectrum of the initial refractive index. In addition, the new indices are not required to be PT-symmetric to allow all-real eigenvalues in their point spectrum. The transition from these results to the time-dependent case is straightforward~\cite{Zel21}, where PT-symmetric structures find interesting applications~\cite{Con19,Con20}.

We have addressed the investigation to obtain guided TE modes in the new waveguiding-structures. This is the reason for which we started from initial refractive indices admitting no leaky modes. In previous works we have studied such possibility by analyzing the resonances of the initial structure~\cite{Raz19,Cru15a,Cru15b}. It is viable to construct the supersymmetric partners using resonances of the initial system~\cite{Fer08}, a technique implemented also in the cosh-like case~\cite{Cev16,Civ19} and for soliton-like models~\cite{Ros20}. However, the transformation of resonances and/or using resonances is elaborated, so it will be analyzed elsewhere. An important point to notice is that although generated from transparent refractive indices, the new structures presented here lack this property as a consequence of the non-Hermiticity (the clear exception is the real-valued case, since it is well known that supersymmetry leaves transparency invariant for such systems). Insights on the matter have been presented by the PT symmetry community and will be considered for future progress of our model.

\appendix
\section{Supersymmetric Finite-Difference Algorithm}
\label{ApA}

\renewcommand{\thesection}{A-\arabic{section}}
\setcounter{section}{0}  

\renewcommand{\theequation}{A-\arabic{equation}}
\setcounter{equation}{0}  

For the sake of completeness, we briefly revisit the generalities of the finite-difference algorithm for higher-order supersymmetry, full details can be consulted in~\cite{Mie00}.

Using the shortcut notation $f_k(x;\epsilon):= f_k(x;\epsilon_1, \epsilon_2, \cdots, \epsilon_{k-1}, \epsilon)$, with $k \geq 1$, the Darboux transformation of an exactly solvable potential $V_{k-1}(x; \epsilon_{k-1})$ produces a new potential $V_k(x;\epsilon)$ in the form
\be
V_k(x;\epsilon) = V_{k-1}(x;\epsilon_{k-1}) + 2\beta'_k(x;\epsilon), \quad k=1,2,\ldots,
\label{darboux}
\ee
where $\beta_k (x; \epsilon)$, usually called the superpotential, is solution of the Riccati equation with the initial potential $V_{k-1}$,
\be
-\beta'_k(x;\epsilon) + \beta_k^2(x;\epsilon) = V_{k-1}(x; \epsilon_{k-1}) - \epsilon,
\label{riccati}
\ee
and $\epsilon$ is a constant to be determined. Although the general solution of (\ref{riccati}) may be found by quadratures~\cite{Inc56}, it is profitable to note that the transformation 
\be
\beta_k (x; \epsilon) = -\frac{d}{dx} \ln u_{(k)} (x; \epsilon)
\label{super}
\ee
linearizes Equation~(\ref{riccati}) by providing the eigenvalue problem 
\be
-u_{(k)}''(x;\epsilon) + V_{k-1} (x; \epsilon_{k-1}) u_{(k)}(x;\epsilon) = \epsilon u_{(k)}(x;\epsilon), \quad k-1,2,\ldots,
\label{uec}
\ee
where $f'(x) = \frac{d}{dx} f(x)$. Thus, the superpotential (\ref{super}) may be constructed from the eigenfunctions $u_{(k)}(x; \epsilon)$ of $V_{k-1}(x; \epsilon_{k-1})$ that belong to the eigenvalue $\epsilon$ (usually called factorization constant). Please note that the `transformation functions' $u_{(k)}$ are just a mathematical tool in the Darboux transformation, so they are not required to be square-integrable in $\operatorname{Dom}V_{k-1}$.

The finite-difference algorithm~\cite{Mie00} states that the solutions of (\ref{riccati}) are the result of a finite-difference operation performed on $\beta_{k-1}$,
\be
\beta_k(x; \epsilon) = -\beta_{k-1}(x;\epsilon_{k-1}) - \frac{\epsilon_{k-1} -\epsilon}{\beta_{k-1}(x,\epsilon_{k-1}) -\beta_{k-1} (x; \epsilon)}.
\label{finite}
\ee
The superpotentials $\beta_k$ constructed at each step automatically solve the Riccati \linebreak\mbox{Equation~(\ref{riccati})} and are linked to the eigenvalue problem (\ref{uec}), which is defined by the potential of the previous step $V_{k-1}$ through the new factorization constant $\epsilon$. In turn, the solutions $\psi_{(k)} (x; \epsilon)$ of the new eigenvalue equation
\be
-\psi_{(k)}''(x;\epsilon) + V_k (x; \epsilon) \psi_{(k)}(x;\epsilon) = \mathcal{E} \psi_{(k)}(x;\epsilon), \quad k=1,2,\ldots,
\label{uec2}
\ee
are easily obtained as the Darboux-deformation of the previous ones:
\be
\mathcal{N}_{(k)}^{-1} \psi_{(k)}(x;\epsilon) = \psi_{(k-1)}'(x;\epsilon_{k-1}) + \beta_k(x;\epsilon) \psi_{(k-1)}(x; \epsilon_{k-1}),
\label{trans}
\ee
where $\mathcal{N}_{(k)}^{-1}$ stands for normalization. 

The breakthrough of the method is the recognition of an additional solution to \linebreak Equation~(\ref{uec2}), which is not included in the transformation (\ref{trans}), given by the expression 
\be
\psi_{(k)}^M (x; \epsilon) = \mathcal{N}_{(k)}^M u^{-1}_{(k)}(x; \epsilon)  =  \mathcal{N}_{(k)}^M \exp \left[\int \beta_k(x; \epsilon) dx \right].
\label{missing}
\ee
The above function was introduced by Mielnik~\cite{Mie84}, it is known as missing state and satisfies~(\ref{uec2}). Thus, if (\ref{missing}) is square-integrable in $\operatorname{Dom}V_k$, then $\epsilon$ must be added to the point spectrum of $V_k$.

\section*{Acknowledgments}

This research has been funded by Consejo Nacional de Ciencia y Tecnolog\'ia (CONACyT), Mexico, Grant Numbers A1-S-24569 and CF19-304307, and Instituto Polit\'ecnico Nacional (IPN), Grant SIP20211204.

A. Romero-Osnaya acknowledges the support from CONACyT through the scholarship 424582.


\end{document}